%

\input harvmac

\input epsf
\ifx\epsfbox\UnDeFiNeD\message{(NO epsf.tex, FIGURES WILL BE
IGNORED)}
\def\figin#1{\vskip2in}
\else\message{(FIGURES WILL BE INCLUDED)}\def\figin#1{#1}\fi
\def\ifig#1#2#3{\xdef#1{fig.~\the\figno}
\goodbreak\topinsert\figin{\centerline{#3}}%
\smallskip\centerline{\vbox{\baselineskip12pt
\advance\hsize by -1truein\noindent{\bf Fig.~\the\figno:} #2}}
\bigskip\endinsert\global\advance\figno by1}


%
%

\noblackbox
%


\def\unlockat{\catcode`\@=11}
\def\lockat{\catcode`\@=12}

\unlockat

\def\newsec#1{\global\advance\secno by1\message{(\the\secno. #1)}
\global\subsecno=0\global\subsubsecno=0\eqnres@t\noindent
{\bf\the\secno. #1}
\writetoca{{\secsym} {#1}}\par\nobreak\medskip\nobreak}
\global\newcount\subsecno \global\subsecno=0
\def\subsec#1{\global\advance\subsecno
by1\message{(\secsym\the\subsecno. #1)}
\ifnum\lastpenalty>9000\else\bigbreak\fi\global\subsubsecno=0
\noindent{\it\secsym\the\subsecno. #1}
\writetoca{\string\quad {\secsym\the\subsecno.} {#1}}
\par\nobreak\medskip\nobreak}
\global\newcount\subsubsecno \global\subsubsecno=0
\def\subsubsec#1{\global\advance\subsubsecno by1
\message{(\secsym\the\subsecno.\the\subsubsecno. #1)}
\ifnum\lastpenalty>9000\else\bigbreak\fi
\noindent\quad{\secsym\the\subsecno.\the\subsubsecno.}{#1}
\writetoca{\string\qquad{\secsym\the\subsecno.\the\subsubsecno.}{#1}}
\par\nobreak\medskip\nobreak}

\def\subsubseclab#1{\DefWarn#1\xdef
#1{\noexpand\hyperref{}{subsubsection}%
{\secsym\the\subsecno.\the\subsubsecno}%
{\secsym\the\subsecno.\the\subsubsecno}}%
\writedef{#1\leftbracket#1}\wrlabeL{#1=#1}}
\lockat

\def\ofo{ { {}_2 \! F_1 }}

\font\manual=manfnt \def\dbend{\lower3.5pt\hbox{\manual\char127}}

\def\IZ{\relax\ifmmode\mathchoice
{\hbox{\cmss Z\kern-.4em Z}}{\hbox{\cmss Z\kern-.4em Z}}
{\lower.9pt\hbox{\cmsss Z\kern-.4em Z}}
{\lower1.2pt\hbox{\cmsss Z\kern-.4em Z}}\else{\cmss Z\kern-.4em
Z}\fi}

\def\p{\partial}


\def\imp{$\Rightarrow$}
\def\IZ{\relax\ifmmode\mathchoice
{\hbox{\cmss Z\kern-.4em Z}}{\hbox{\cmss Z\kern-.4em Z}}
{\lower.9pt\hbox{\cmsss Z\kern-.4em Z}}
{\lower1.2pt\hbox{\cmsss Z\kern-.4em Z}}\else{\cmss Z\kern-.4em
Z}\fi}
\def\IB{\relax{\rm I\kern-.18em B}}
\def\IC{{\relax\hbox{$\inbar\kern-.3em{\rm C}$}}}
\def\ID{\relax{\rm I\kern-.18em D}}
\def\IE{\relax{\rm I\kern-.18em E}}
\def\IF{\relax{\rm I\kern-.18em F}}
\def\IG{\relax\hbox{$\inbar\kern-.3em{\rm G}$}}
\def\IGa{\relax\hbox{${\rm I}\kern-.18em\Gamma$}}
\def\IH{\relax{\rm I\kern-.18em H}}
\def\II{\relax{\rm I\kern-.18em I}}
\def\IK{\relax{\rm I\kern-.18em K}}
\def\IP{\relax{\rm I\kern-.18em P}}
\def\IQ{\relax\hbox{$\inbar\kern-.3em{\rm Q}$}}

\def\inbar{\,\vrule height1.5ex width.4pt depth0pt}

\def\p{\partial}

\font\cmss=cmss10 \font\cmsss=cmss10 at 7pt
\def\IR{\relax{\rm I\kern-.18em R}}

%
%

\def\makeblankbox#1#2{\hbox{\lower\dp0\vbox{\hidehrule{#1}{#2}%
   \kern -#1
   \hbox to \wd0{\hidevrule{#1}{#2}%
      \raise\ht0\vbox to #1{}
      \lower\dp0\vtop to #1{}
      \hfil\hidevrule{#2}{#1}}%
   \kern-#1\hidehrule{#2}{#1}}}%
}%
\def\hidehrule#1#2{\kern-#1\hrule height#1 depth#2 \kern-#2}%
\def\hidevrule#1#2{\kern-#1{\dimen0=#1\advance\dimen0 by #2\vrule
    width\dimen0}\kern-#2}%
\def\openbox{\ht0=1.2mm \dp0=1.2mm \wd0=2.4mm  \raise 2.75pt
\makeblankbox {.25pt} {.25pt}  }

\def\bun#1/#2{\leavevmode
   \kern.1em \raise .5ex \hbox{\the\scriptfont0 #1}%
   \kern-.1em $/$%
   \kern-.15em \lower .25ex \hbox{\the\scriptfont0 #2}%
}

\def\opensquare{\ht0=3.4mm \dp0=3.4mm \wd0=6.8mm  \raise 2.7pt
\makeblankbox {.25pt} {.25pt}  }


\def\sector#1#2{\ {\scriptstyle #1}\hskip 1mm
\mathop{\opensquare}\limits_{\lower 1mm\hbox{$\scriptstyle#2$}}\hskip 1mm}

\def\tsector#1#2{\ {\scriptstyle #1}\hskip 1mm
\mathop{\opensquare}\limits_{\lower 1mm\hbox{$\scriptstyle#2$}}^\sim\hskip 1mm}


\def\inbar{\,\vrule height1.5ex width.4pt depth0pt}

\def\p{\partial}

\font\cmss=cmss10 \font\cmsss=cmss10 at 7pt
\def\IR{\relax{\rm I\kern-.18em R}}


\def\frac#1#2{{#1\over#2}}

\def\inbar{\,\vrule height1.5ex width.4pt depth0pt}
\def\IC{\relax\hbox{$\inbar\kern-.3em{\rm C}$}}
\def\IR{\relax{\rm I\kern-.18em R}}
\def\IP{\relax{\rm I\kern-.18em P}}

%
%
\catcode`\@=11
\def\slash#1{\mathord{\mathpalette\c@ncel{#1}}}
\overfullrule=0pt

\def\II{{\cal I}}

\def\NN{{\cal N}}
\def\OO{{\cal O}}

\def\underrel#1\over#2{\mathrel{\mathop{\kern\z@#1}\limits_{#2}}}

\catcode`\@=12


%

\def\cosh{{\rm cosh}}


\def\ra{{\rightarrow}}

\def\ra {{\rightarrow}}

\def\w{{\bf w}}
\def\q{{\bf q}}
\def\F{{\cal F}}

%



\lref\forster{D.~Forster,  {\it ``Hydrodynamic Fluctuations,
Broken Symmetry, and Correlation Functions''}, W.~A.~Benjamin,
Inc., Reading, Massachusetts, 1975.}

\lref\landaulp{E.~M.~Lifshits and L.~P.~Pitaevskii,  {\it
``Statistical physics, Part 2''}, Pergamon Press, New York, 1980.}

\lref\stanley{H.~E.~Stanley, {\it "Introduction to phase
transitions and critical phenomena"}, Clarendon Press, Oxford,
1971.}

\lref\landau{L.~D.~Landau and E.~M.~Lifshits,  {\it
``Fluid Mechanics''},
Pergamon Press, New York, 1987, 2nd ed.}

\lref\aps{
A.~Adams, J.~Polchinski and E.~Silverstein,
``Don't panic! Closed string tachyons in ALE space-times,''
JHEP {\bf 0110}, 029 (2001)
[arXiv:hep-th/0108075].
}

\lref\GiveonRW{
  A.~Giveon, A.~Konechny, E.~Rabinovici and A.~Sever,
  ``On thermodynamical properties of some coset CFT backgrounds,''
  JHEP {\bf 0407}, 076 (2004)
  [arXiv:hep-th/0406131].
}

\lref\KS{
  D.~Kutasov and D.~A.~Sahakyan,
  ``Comments on the thermodynamics of little string theory,''
  JHEP {\bf 0102}, 021 (2001)
  [arXiv:hep-th/0012258].
}

\lref\GKb{
  A.~Giveon and D.~Kutasov,
  ``Comments on double scaled little string theory,''
  JHEP {\bf 0001}, 023 (2000)
  [arXiv:hep-th/9911039].
}

\lref\GKa{
  A.~Giveon and D.~Kutasov,
  ``Little string theory in a double scaling limit,''
  JHEP {\bf 9910}, 034 (1999)
  [arXiv:hep-th/9909110].
}

\lref\AFKS{
  O.~Aharony, B.~Fiol, D.~Kutasov and D.~A.~Sahakyan,
  ``Little string theory and heterotic/type II duality,''
  Nucl.\ Phys.\ B {\bf 679}, 3 (2004)
  [arXiv:hep-th/0310197].
}

\lref\GKPS{
  A.~Giveon, A.~Konechny, A.~Pakman and A.~Sever,
  ``Type 0 strings in a 2-d black hole,''
  JHEP {\bf 0310}, 025 (2003)
  [arXiv:hep-th/0309056].
}

\lref\NarayanDR{
  K.~Narayan and M.~Rangamani,
  ``Hot little string correlators: A view from supergravity,''
  JHEP {\bf 0108}, 054 (2001)
  [arXiv:hep-th/0107111].
}

\lref\DeBoerDD{
  P.~A.~DeBoer and M.~Rozali,
  ``Thermal correlators in little string theory,''
  Phys.\ Rev.\ D {\bf 67}, 086009 (2003)
  [arXiv:hep-th/0301059].
}

\lref\HorowitzCD{
  G.~T.~Horowitz and A.~Strominger,
  ``Black strings and P-branes,''
  Nucl.\ Phys.\ B {\bf 360}, 197 (1991).
}

\lref\MaldacenaCG{
  J.~M.~Maldacena and A.~Strominger,
  ``Semiclassical decay of near-extremal fivebranes,''
  JHEP {\bf 9712}, 008 (1997)
  [arXiv:hep-th/9710014].
}

\lref\SonSD{
  D.~T.~Son and A.~O.~Starinets,
  ``Minkowski-space correlators in AdS/CFT correspondence: Recipe and
  applications,''
  JHEP {\bf 0209}, 042 (2002)
  [arXiv:hep-th/0205051].
}

\lref\PSS{
  G.~Policastro, D.~T.~Son and A.~O.~Starinets,
  ``From AdS/CFT correspondence to hydrodynamics,''
  JHEP {\bf 0209}, 043 (2002)
  [arXiv:hep-th/0205052].
}

\lref\KST{
  P.~K.~Kovtun and A.~O.~Starinets,
  ``Quasinormal modes and holography,''
  arXiv:hep-th/0506184.
}

%
%
%

\lref\HarmarkHW{
  T.~Harmark and N.~A.~Obers,
  ``Hagedorn behaviour of little string theory from string corrections to
  NS5-branes,''
  Phys.\ Lett.\ B {\bf 485}, 285 (2000)
  [arXiv:hep-th/0005021].
}

\lref\BerkoozMZ{
  M.~Berkooz and M.~Rozali,
  ``Near Hagedorn dynamics of NS fivebranes, or a new universality class  of
  coiled strings,''
  JHEP {\bf 0005}, 040 (2000)
  [arXiv:hep-th/0005047].
}

\lref\KSSa{
  P.~Kovtun, D.~T.~Son and A.~O.~Starinets,
  ``Holography and hydrodynamics: Diffusion on stretched horizons,''
  JHEP {\bf 0310}, 064 (2003)
  [arXiv:hep-th/0309213].
}

\lref\KSSb{
  P.~Kovtun, D.~T.~Son and A.~O.~Starinets,
  ``Viscosity in strongly interacting quantum field theories from black hole
  physics,''
  Phys.\ Rev.\ Lett.\  {\bf 94}, 111601 (2005)
  [arXiv:hep-th/0405231].
}

\lref\StarinetsBR{
  A.~O.~Starinets,
  ``Quasinormal modes of near extremal black branes,''
  Phys.\ Rev.\ D {\bf 66}, 124013 (2002)
  [arXiv:hep-th/0207133].
}

\lref\AharonyTT{
  O.~Aharony and T.~Banks,
  ``Note on the quantum mechanics of M theory,''
  JHEP {\bf 9903}, 016 (1999)
  [arXiv:hep-th/9812237].
}

\lref\BerkoozMZ{
  M.~Berkooz and M.~Rozali,
  ``Near Hagedorn dynamics of NS fivebranes, or a new universality class  of
  coiled strings,''
  JHEP {\bf 0005}, 040 (2000)
  [arXiv:hep-th/0005047].
}

\lref\RangamaniIR{
  M.~Rangamani,
  ``Little string thermodynamics,''
  JHEP {\bf 0106}, 042 (2001)
  [arXiv:hep-th/0104125].
}

\lref\BuchelDG{
  A.~Buchel,
  ``On the thermodynamic instability of LST,''
  arXiv:hep-th/0107102.
}

\lref\Seiberg{
  N.~Seiberg,
  ``New theories in six dimensions and matrix description of M-theory on  T**5
  and T**5/Z(2),''
  Phys.\ Lett.\ B {\bf 408}, 98 (1997)
  [arXiv:hep-th/9705221].
}

\lref\bbs{
  P.~Benincasa, A.~Buchel and A.~O.~Starinets,
  ``Sound waves in strongly coupled non-conformal gauge theory plasma,''
  arXiv:hep-th/0507026.
}

%
%

\lref\BuchelTZ{
  A.~Buchel and J.~T.~Liu,
  ``Universality of the shear viscosity in supergravity,''
  Phys.\ Rev.\ Lett.\  {\bf 93}, 090602 (2004)
  [arXiv:hep-th/0311175].
}

\lref\BuchelQQ{
  A.~Buchel,
  ``On universality of stress-energy tensor correlation functions in
  Phys.\ Lett.\ B {\bf 609}, 392 (2005)
  [arXiv:hep-th/0408095].
}

\lref\HerzogFN{
  C.~P.~Herzog,
  ``The hydrodynamics of M-theory,''
  JHEP {\bf 0212}, 026 (2002)
  [arXiv:hep-th/0210126].
}

\lref\AharonyTT{
  O.~Aharony and T.~Banks,
  ``Note on the quantum mechanics of M theory,''
  JHEP {\bf 9903}, 016 (1999)
  [arXiv:hep-th/9812237].
}

\lref\RangamaniIR{
  M.~Rangamani,
  ``Little string thermodynamics,''
  JHEP {\bf 0106}, 042 (2001)
  [arXiv:hep-th/0104125].
}

\lref\BuchelDG{
  A.~Buchel,
  ``On the thermodynamic instability of LST,''
  arXiv:hep-th/0107102.
}

\lref\AharonyXN{
  O.~Aharony, A.~Giveon and D.~Kutasov,
  ``LSZ in LST,''
  Nucl.\ Phys.\ B {\bf 691}, 3 (2004)
  [arXiv:hep-th/0404016].
}

\lref\GiveonZM{
  A.~Giveon, D.~Kutasov and O.~Pelc,
  ``Holography for non-critical superstrings,''
  JHEP {\bf 9910}, 035 (1999)
  [arXiv:hep-th/9907178].
}

\lref\OoguriWJ{
  H.~Ooguri and C.~Vafa,
  ``Two-Dimensional Black Hole and Singularities of CY Manifolds,''
  Nucl.\ Phys.\ B {\bf 463}, 55 (1996)
  [arXiv:hep-th/9511164].
}

\lref\KutasovTE{
  D.~Kutasov,
  ``Orbifolds and Solitons,''
  Phys.\ Lett.\ B {\bf 383}, 48 (1996)
  [arXiv:hep-th/9512145].
}

\lref\AharonyUB{
  O.~Aharony, M.~Berkooz, D.~Kutasov and N.~Seiberg,
  ``Linear dilatons, NS5-branes and holography,''
  JHEP {\bf 9810}, 004 (1998)
  [arXiv:hep-th/9808149].
}

\lref\MinwallaXI{
  S.~Minwalla and N.~Seiberg,
  ``Comments on the IIA NS5-brane,''
  JHEP {\bf 9906}, 007 (1999)
  [arXiv:hep-th/9904142].
}

\lref\PeetWN{
  A.~W.~Peet and J.~Polchinski,
  ``UV/IR relations in AdS dynamics,''
  Phys.\ Rev.\ D {\bf 59}, 065011 (1999)
  [arXiv:hep-th/9809022].
}

\lref\KapustinCI{
  A.~Kapustin,
  ``On the universality class of little string theories,''
  Phys.\ Rev.\ D {\bf 63}, 086005 (2001)
  [arXiv:hep-th/9912044].
}

\lref\AtickSI{
  J.~J.~Atick and E.~Witten,
  ``The Hagedorn Transition And The Number Of Degrees Of Freedom Of String
  Theory,''
  Nucl.\ Phys.\ B {\bf 310}, 291 (1988).
}

\lref\BuchelDI{
  A.~Buchel, J.~T.~Liu and A.~O.~Starinets,
  ``Coupling constant dependence of the shear viscosity in N = 4 supersymmetric
  Yang-Mills theory,''
  Nucl.\ Phys.\ B {\bf 707}, 56 (2005)
  [arXiv:hep-th/0406264].
}

\lref\KlemmBJ{
  A.~Klemm, W.~Lerche, P.~Mayr, C.~Vafa and N.~P.~Warner,
  ``Self-Dual Strings and N=2 Supersymmetric Field Theory,''
  Nucl.\ Phys.\ B {\bf 477}, 746 (1996)
  [arXiv:hep-th/9604034].
}

\lref\MaldacenaHW{
  J.~M.~Maldacena and H.~Ooguri,
  ``Strings in AdS(3) and SL(2,R) WZW model. I,''
  J.\ Math.\ Phys.\  {\bf 42}, 2929 (2001)
  [arXiv:hep-th/0001053].
}

\lref\ArgyresXN{
  P.~C.~Argyres, M.~Ronen Plesser, N.~Seiberg and E.~Witten,
  ``New N=2 Superconformal Field Theories in Four Dimensions,''
  Nucl.\ Phys.\ B {\bf 461}, 71 (1996)
  [arXiv:hep-th/9511154].
}

\lref\ArgyresJJ{
  P.~C.~Argyres and M.~R.~Douglas,
  ``New phenomena in SU(3) supersymmetric gauge theory,''
  Nucl.\ Phys.\ B {\bf 448}, 93 (1995)
  [arXiv:hep-th/9505062].
}

\lref\EguchiVU{
  T.~Eguchi, K.~Hori, K.~Ito and S.~K.~Yang,
  ``Study of $N=2$ Superconformal Field Theories in $4$ Dimensions,''
  Nucl.\ Phys.\ B {\bf 471}, 430 (1996)
  [arXiv:hep-th/9603002].
}


%
\Title{\vbox{\baselineskip12pt
\hbox{hep-th/0506144}
}}
{\vbox{\centerline{The Silence of the Little Strings}}}
\centerline{Andrei Parnachev$^1$ and Andrei Starinets$^2$}
\bigskip
\centerline{$^1${\it Department of Physics, Rutgers University}}
\centerline{\it Piscataway, NJ 08854-8019, USA}
\centerline{$^2$ {\it Perimeter Institute for Theoretical
Physics}} \centerline{\it Waterloo, ON, N2L 2Y5, Canada}
\bigskip
 \vskip.1in \vskip.1in \centerline{\bf Abstract}
\noindent We study the hydrodynamics of the high-energy phase of
Little String Theory.
The poles of the retarded two-point function of the stress energy tensor
 contain information about the speed of sound and the kinetic
coefficients, such as shear and bulk  viscosity.
We compute this two-point function in the dual string theory
and analytically continue it to Lorentzian signature.
We perform an independent check of our results
by the Lorentzian supergravity calculation in the background of
 non-extremal NS5-branes.
The speed of sound vanishes at the Hagedorn temperature. The ratio
of shear viscosity to entropy density is equal to the universal
value $1/4\pi$ and does not receive $\alpha'$  corrections. The
ratio of bulk viscosity to entropy density equals  $1/10\pi$. We
also compute the $R$-charge diffusion constant. In addition to the
hydrodynamic singularities, the correlators have an infinite
series of finite-gap
 poles, and a massless pole with zero attenuation.
%
%
%



\Date{September 28, 2005}



\newsec{Introduction and summary}
Little String Theory (LST) is a nonlocal theory without gravity
which can be defined as the theory of NS5-branes in the limit of
vanishing string coupling \Seiberg. In this limit bulk modes
decouple, but the theory on the five-branes remains nontrivial. An
alternative definition involves formulating string theory on a
Calabi-Yau space and going to a singular point in the moduli space
of the Calabi-Yau \GiveonZM.
These two formulations are related by T-duality \refs{\OoguriWJ,\KutasovTE}.
In both definitions one can make the theory amenable to a perturbative
description: in the five-brane language this involves separating branes,
while in the Calabi-Yau picture one needs to resolve the singularity and to 
take
a certain weak coupling limit.

A collection  of non-extremal NS5 branes describes a
high-energy phase of LST. Thermodynamics of this system has
been studied in
\refs{\MaldacenaCG\AharonyTT\HarmarkHW\BerkoozMZ\KS\RangamaniIR\BuchelDG\NarayanDR\DeBoerDD-\AharonyXN}.
Classically, the theory has a Hagedorn density of states and the
temperature is fixed at the Hagedorn value $T_H$ which depends on the
number of five-branes $k$, but is independent of the energy
density. This theory has an exact CFT description. When string
loop corrections are included, the temperature of the system
may differ from $T_H$. The one-loop calculation
shows that the specific heat is negative in the regime of
high energy density \KS.
The absolute value of the specific heat diverges
as the temperature approaches $T_H$ from above. Hence, this phase of LST
is unstable, similar to a Schwarzschild black hole or a small
black hole in AdS space. One would expect a more conventional
phase to appear at low energy densities, where the theory on the
NS5-branes reduces to (1,1) superconformal Yang-Mills for IIB or
(2,0) superconformal theory for IIA string theory. (This regime is
not accessible in the dual string theory which becomes strongly
coupled \AharonyUB.)

Temperature as a function of energy for the low- and high-energy
phases of LST is shown schematically in Figure 1. In \KS\ it has
been argued that the Hagedorn temperature $T_H$ is reached from
below at a finite energy $E_*$. In this paper, we study
hydrodynamics of the high-temperature phase of LST. Our analysis
corresponds to $E\rightarrow \infty$, where $T$ approaches $T_H$
from above.
%
%
%
\ifig\frequf{Temperature as a function of energy in LST. The left
branch assumes a dependence similar to that of the six-dimensional
Yang-Mills theory in the infrared, while the right branch is the
high-energy phase above the Hagedorn temperature. The right branch
has negative specific heat.
Our analysis
 corresponds
 to $E\rightarrow \infty$, where $T$ approaches $T_H$ from above.}
{\epsfxsize=9.5cm \epsfysize=5.5cm \epsfbox{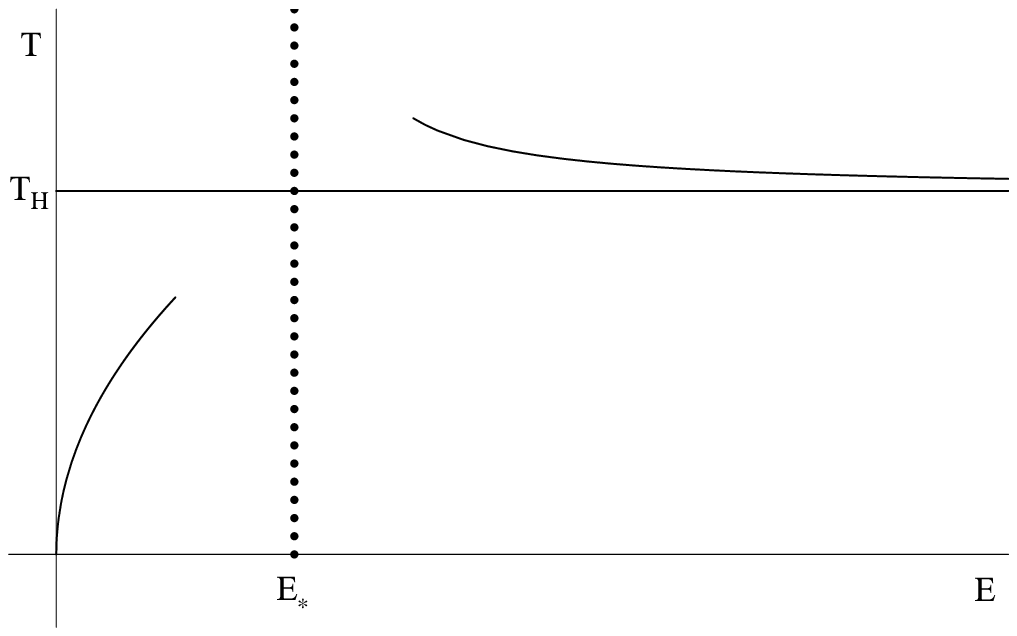}}
%

The hydrodynamics of black branes has been considered in
\refs{\PSS\HerzogFN-\KSSa}. More precisely, one can determine the
speed of sound and the kinetic coefficients, 
such as shear and bulk viscosity, 
 for the theories whose dual description (in a
certain regime)  is given by a supergravity background involving
black branes. The transport coefficients can be found by taking
the hydrodynamic limit in thermal two-point functions of the
operators corresponding to conserved currents (e.g. stress-energy
tensor), or, equivalently, by identifying gapless quasinormal
frequencies of the supergravity background \refs{\SonSD, \StarinetsBR, \KST}.
Remarkably, for a large class of theories in the regime described
by supergravity duals, the ratio of shear viscosity to entropy
density has the universal value of $1/4\pi$
\refs{\KSSa,\KSSb\BuchelTZ -\BuchelQQ}.
 Computing bulk viscosity is a more arduous task, since
in the supergravity description it involves considering diagonal
components of the metric perturbation. If bulk viscosity is
non-zero, the diagonal components will couple to fluctuations of
the fields in the system responsible for breaking the conformal
invariance (e.g. fluctuations of the dilaton). Thus it is not
surprising that in computing bulk viscosity even for a relatively
simple non-conformal background one is compelled to resort to
numerical methods \bbs. However, we shall see that in the
high-temperature phase of LST in the limit $T\rightarrow T_H$ the
ratio of bulk viscosity to entropy
 density can be computed analytically.
Moreover, the existence of an exact CFT description allows us to
compute transport coefficients to all orders in $\alpha'$.

We determine the transport coefficients by two independent methods:
first, by computing the two-point functions of the stress-energy tensor
 and the $R$-currents using the exact CFT description, and then by
finding the  quasinormal spectrum of the non-extremal NS5-brane
background. We find a complete agreement between two approaches.
We compare our results with the analysis of linearized hydrodynamics.
Another feature of LST is its non-locality, whose scale is set by
$\sqrt{k} l_s=\sqrt{k \alpha'}$. This should not be of
significance for the hydrodynamic regime,
as the wavelength
of hydrodynamic excitations is much larger then
$1/T_H\sim \sqrt{k} l_s$.

In summary, we find that when the temperature approaches $T_H$
 from above, the speed of sound vanishes, the ratio of shear
viscosity to entropy density
 is equal to the universal value $1/ 4\pi$, the ratio of bulk
 viscosity to
entropy density equals to $1/ 10 \pi$, and
 the $R$-charge diffusion constant is $1/4 \pi T_H$.


%
%
%

The paper is organized as follows. In Section 2 we review the
thermodynamics of LST, including the first correction to classical
thermodynamics coming from the loop expansion in string theory.
In the high energy limit, the pressure behaves as $P\sim\log E$,
which implies  that the speed of sound $v_s \sim 1/\sqrt{E}$ vanishes at $T_H$.
%
%
While this might seem unusual, we note that
in models describing 
  conventional systems, vanishing or a sharp decrease of the
speed of sound is related to
 a  phase transition.
Indeed, the speed of sound  is given by $v_s =(1/\rho
\kappa)^{-1/2}$, where $\kappa$ is the compressibility and $\rho$
is the equilibrium mass density of the system. At a liquid-gas
critical point, the compressibility diverges as $\kappa \sim
(T-T_c)^{-\gamma}$, where $\gamma \approx 1.3$ \stanley, which implies
$v_s\ra 0$ (see Fig.~2).
One can also show that the speed of sound decreases sharply when
the waves propagate through a two-phase medium (e.g. a liquid with
bubbles of gas in it) near the transition point \landau\ .
\ifig\frequf{Isothermal speed of sound $v_s =\sqrt{(\partial
P/\partial n)_T}$ as a function of density in the van der Waals
model of liquid-gas phase transition. } {\epsfxsize=9.5cm
\epsfysize=5.5cm \epsfbox{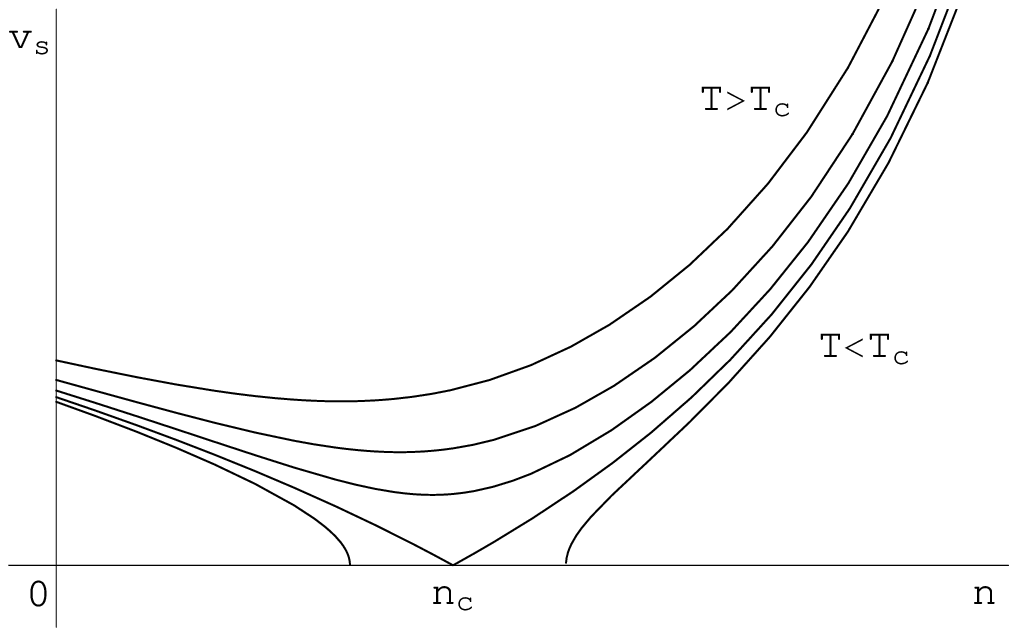}}

In Section 3 we consider hydrodynamics of LST. In the limit of
vanishing $v_s$, the propagating mode effectively becomes a
diffusive one, due to non-zero attenuation. Moreover, for a
certain value of the ratio of bulk to shear viscosity, one of the
components of the stress-energy tensor in the hydrodynamic
constitutive relation decouples from the rest.
 On the level of the stress-energy tensor  two-point functions
 this means that while all the correlators in the sound channel
have the same pole in the hydrodynamic regime, the correlator
corresponding to the decoupled mode has none. (This is exactly
what we observe when computing the LST correlators in string
theory and supergravity.)

In Sections 4, 5 and 6 we compute the two-point
function of the stress energy tensor $T_{\mu\nu}$. The computation
is done in the dual string theory involving  the Euclidean
$SL(2)/U(1)$ (cigar) background, and the amplitudes are then
analytically continued to the Lorentzian signature. The two-point
function of the components of $T_{\mu\nu}$ corresponding to the shear mode
exhibits a hydrodynamic pole at $\omega = -i q^2/4\pi T$. This
implies that the shear viscosity to entropy ratio is equal to the
universal value $1/4\pi$.
Our result is exact to all orders in $\alpha'$. On the other hand,
it is only valid at the Hagedorn temperature. Extending it to
other temperatures requires the analysis of string loop corrections to
the two-point functions.
The Green's function of the stress-energy tensor components 
corresponding to the sound mode is also
computed. It turns
 out that the correlator has a double pole at  $\omega = -i q^2/4\pi T$
which
 is consistent with the observation that the
speed of sound vanishes at $T_H$ as well as the ratio of bulk
 viscosity to entropy density reported above.

In Section 7 we verify the string theory results by computing the
quasinormal spectrum of the non-extremal NS5-brane background.
 Interestingly, the poles observed in supergravity agree
with the string theory results exactly and do not receive $1/k$
corrections.\foot{ This has been observed for the scalar mode in
\NarayanDR.} There are additional poles in string theory which are
not visible in supergravity, but these do not appear in the
hydrodynamic regime.
We discuss our results in Section 8.

%
%

\newsec{Review of Little String Theory Thermodynamics}
We start by reviewing the thermodynamics of LST, closely following the
 presentation in \KS.
The supergravity solution for the $k$ coincident non-extremal
NS5-branes in the string frame is \refs{\HorowitzCD,\MaldacenaCG}
 \eqn\nensfive{ ds^2 = - f(r) dt^2 + dx_5^2
   +  A(r)
    \left({dr^2\over f(r)} + r^2 d\Omega_3^2\right)\,,}
\eqn\ldilo{ e^{2\Phi}=g_s^2 A(r)\,,} \eqn\formo{ H_3 = 2 L
\sqrt{k \alpha ' (r_0^2 + k \alpha ')} \; \epsilon_3\,,}
 \eqn\newfo{f(r) = 1- {r_0^2\over r^2}\,,}
  \eqn\newao{ A(r) = 1 + {k \alpha '  \over r^2}\,,}
 where $r_0$ is the location of
the horizon,
$dx_5^2$ denotes the metric along the five-brane flat
directions\foot{We denote the spatial
 coordinates along the five flat directions by $z, x^a$, $a=1,2,3,4$,
singling out one of the directions, $z$, which we orient along
the spatial momentum.}, $d\Omega_3$ is the metric and $\epsilon_3$ is the
volume form of the unit three-sphere.
%
%
The energy above extremality, per unit volume,
 for the solution \nensfive -- \newao\ is
\eqn\euv{ \epsilon\equiv{E\over V_5}={1\over(2\pi)^5\alpha'^3 }\,
\mu,\;\;\;\;\; \qquad
          \mu\equiv {r_0^2\over g_s^2 \alpha'}\,.  }
The near-horizon Euclidean geometry is obtained by Wick-rotating
via $t= -i\tau$ and taking $r_0, g_s\ra 0$, keeping the quantity
$\mu$ fixed: \eqn\nhnsfive{ ds^2= k\alpha'
\left(d\phi^2+\tanh^2\phi d\tau^2+d\Omega_3^2\right)+dx_5^2\,,  }
\eqn\nhldil{ e^{2\Phi}={k\over \mu\cosh^2\phi}  \,.  }
The absence of the conical singularity at $\phi=0$ requires $\tau$ to be
$2\pi$-periodic.
The inverse temperature is equal to the
circumference of the temporal circle in \nhnsfive,
\eqn\betah{  \beta_H=2\pi \sqrt{k\alpha'}\,.   }
Strings propagating in the background \nhnsfive, \nhldil\
are described by an exact conformal field theory.
We  review some details of that theory in Section 4.

In the gravity approximation, the inverse temperature is
independent of the energy density. As
 \eqn\betas{  \beta={\p S\over
\p E}=\beta_H  \,,  }
 the entropy is proportional to the energy,
\eqn\hds{  S=\beta_H E\,.}
In the microcanonical ensemble,
this corresponds to a Hagedorn density of states $\rho(E)\sim e^{\beta_H E}$.
When string loop corrections are taken
into account, the density of states is modified according to
\eqn\hdsa{ \rho(E)\sim E^\alpha e^{\beta_H
E} \left[1+\OO\left({1\over E}\right)\right]\,.  } The coefficient
$\alpha$ in \hdsa\ has been computed in \KS, and $\alpha +1$ was
found to be {\it negative}. This has important implications for
the phase structure of LST. The relation \eqn\betaerel{ \beta=\p
S(E)/\p E }
 together with \hdsa\ leads to
the following energy-temperature relation
\eqn\et{  \beta-\beta_H={\alpha\over E}+\OO\left({1\over E^2}\right) \,.  }
Thus for temperatures slightly above the Hagedorn temperature
the energy is given by
\eqn\eaboveht{  E={\alpha\over\beta-\beta_H}
\Biggl[1+\OO(\beta-\beta_H)\Biggr] \,.  }
In this regime, one can perform
 consistent perturbative expansion
in powers of $\beta-\beta_H$ or, equivalently, in powers of $1/E$.
This is the type of expansion we will be interested in.
As discussed below, this corresponds to the genus expansion in the
dual string theory.

When the temperature is slightly below the Hagedorn temperature, 
Eq.~\betaerel\
implies that one has to compute $S(E)$ to all orders in perturbation
theory, and possibly to include non-perturbative corrections.
A generic function $S(E)$ would then mean that the Hagedorn temperature
is reached from below at finite energy.

 Eq.~\hdsa\ implies that
the free energy $\F$ of LST is determined by
\eqn\fe{   -\beta\F=S-\beta E\simeq-(\alpha+1)\log(\beta-\beta_H)\simeq
                    (\alpha+1)\log E \,.  }
In the second equality we used Eq.~\eaboveht. The leading term in the
free energy, which is proportional to energy, vanishes due to Eq.~\hds.
 The
string theory partition function is related to the free energy of
LST via \eqn\pffe{ Z_{string} =  -\beta \F\,.     } The 
genus zero string
partition function is proportional to  energy, \eqn\gzero{ e^{-2\Phi_0}
Z_0={\mu\over k} Z_0 \sim {\epsilon\over k} Z_0 \,,  } but, as
explained in \KS, $Z_0$ vanishes. Hence, to compute the first
non-trivial term in the free energy, one must compute the string
partition function on the torus. This partition function
 is proportional to  $\log E$.
The computation was done in \KS, where the coefficient
$\alpha$ was found to be
\eqn\alphaval{
\alpha = -1 - a_1 V_5\,,  } where $a_1$ is a positive number which
scales as $(k\alpha')^{-5/2}$ \KS .
From \fe\ it follows that the pressure is proportional 
to the logarithm of energy,
$P=-\partial {\cal F}/\partial V_5 \sim a_1 \log E$, and thus the speed of
 sound given by
\eqn\sound{v_s = \sqrt{ {\partial P\over \partial E}} \sim {1\over \sqrt{E}}\,}
vanishes at $T=T_H$.

\newsec{Hydrodynamics of Little String Theory}

Hydrodynamics is an effective theory describing time evolution
of the densities of conserved charges
in the regime of long
wavelengths, i.e. at a scale $l$ such that
\eqn\condi{l_{micro}\ll l \ll L\,,}
where $l_{micro}$ is a characteristic scale of microscopic
processes in the system (e.g. a correlation length), and $L$ is a
typical size of the system.
The hydrodynamic description becomes unreliable when the inequality
\condi\ is not satisfied. 
For example, 
Schwarzschild black holes do not seem to correspond to
any hydrodynamic regime in a (hypothetical) holographically dual description.
Indeed, in that case the characteristic microscopic scale 
(thermal wavelength) is of order $l_{micro}\sim 1/T$, while 
the size of the system (Schwarzschild radius) is $L\sim 1/T$.
%

To derive the dispersion relations for the shear and the sound
modes, consider small deviations from equilibrium
$T_{\mu\nu}=\langle T_{\mu\nu} \rangle
+ \tilde{T}_{\mu\nu}(t,x)$
in
 the stress-energy tensor of a theory in a $D+1$ dimensional Minkowski space.
 The equations of linearized hydrodynamics
follow from the conservation law $\partial_{\mu} T^{\mu\nu}=0$,
\eqn\hrrt{\eqalign{&\partial_0 \tilde{T}^{00} +  \partial_i
\tilde{T}^{0i} =0\,,\cr &\partial_0  \tilde{T}^{0i} + \partial_j
\tilde{T}^{ij} =0\,,\cr} } together with the constitutive
relations which express all components $\tilde{T}_{\mu\nu}$ in
terms of fluctuations  $\tilde{T}^{00}$,  $\tilde{T}^{0i}$ of the
densities of conserved charges (energy and momentum):
\eqn\hbbc{T^{00} = \epsilon + \tilde{T}^{00}\,,} \eqn\hbbc{T^{ij}
= \delta^{ij} \left( P + {\partial P \over \partial \epsilon}\,
\tilde{T}^{00}\right) - {1\over \epsilon + P} \left[ \eta \left(
\partial_i T^{0j} +
\partial_j T^{0i} - {2\over D}  \delta^{ij} \partial_k T^{0k}\right)
+ \zeta  \delta^{ij} \partial_k T^{0k}\right]\,,}
where $\epsilon = <T^{00}>$, $\epsilon$ and $P$ are the equilibrium energy density and pressure, $\eta$ and $\zeta$ are the shear and bulk viscosities, respectively.
Assuming the coordinate dependence of the variables in 
Eq.~\hrrt\ to be of the 
form
$\propto e^{-i \omega t + i q z}$, we find that  the system \hrrt\ has two types of eigenmodes - the shear mode
with the dispersion relation
\eqn\hbbs{\omega = -  {i \eta \over \epsilon + P}\, q^2 =
 -  {i \eta \over s T}\, q^2\,}
and the sound mode whose dispersion relation is determined by the equation
\eqn\hbbx{\omega^2  + i\, \Gamma \, \omega \, q^2
 - v_s^2 q^2=0\,,}
where $v_s = \left(\partial P/\partial \epsilon\right)^{1/2}$ is the speed of sound
and \eqn\hbbxc{\Gamma = {1\over \epsilon + P} \left[ \zeta +
\left( 2-{2\over D} \right)\, \eta \right] } is the damping
constant. For nonvanishing speed of sound the dispersion relation
is \eqn\hbbz{\omega = \pm v_s q -  {i \Gamma \over 2}\, q^2 +
\cdots\,,} where ellipses denote terms suppressed for $q
\Gamma/v_s \ll 1$. However, if $v_s=0$, we find only one
nontrivial solution,
 \eqn\hbbsx{\omega = - i \Gamma \, q^2\,.}

The dispersion relations for the shear and the sound wave modes appear as
the poles of the retarded Green's functions of the stress-energy tensor
 \eqn\rgf{
G_{\mu\nu,\rho\sigma}(\omega ,q)
=-i\int d t d^D x e^{- i \omega t + i q z}\, \theta(t)
        \langle\left[T_{\mu \nu}(t, {\bf x}),T_{\rho \sigma}(0)\right]
\rangle\,. } where ${\bf x}=(x^a,z),\; a=1,\ldots,4$ and the
spatial momentum is chosen along the $z$ direction. In the
hydrodynamic limit $\omega/T\ll1$, $q/T\ll 1$ the correlators are
expected to  have the following pole structure \landaulp ,
\forster\ (see also \KST ) :


$\bullet$ Each of the shear mode correlators $G_{zx^a,zx^a}$,
$G_{tx^a,tx^a}$,  $G_{tx^a,zx^a}$, where $x^a\neq z$,
 has a pole at $\omega$ given by \hbbs\ .

$\bullet$  The scalar mode correlators  $G_{x^ax^b,x^ax^b}$, where
$x^a\neq z$, $a\neq b$, do not exhibit hydrodynamic poles.

$\bullet$ The correlators of the sound mode,  $G_{tt,tt}$,
$G_{zz,zz}$,  $G_{tz,tz}$, all have poles at  $\omega$ given by
\hbbz\ , or, if $v_s=0$, by \hbbsx\ . The correlator  $G_{x^a x^a,
x^a x^a}$,  where $x^a\neq z$, belongs to the same family, unless
 \eqn\rgfs{v_s = 0 \,, \;\;\; \zeta = {2 \over D}\, \eta\,,}
in which case the corresponding mode decouples from the sound wave mode,
as follows from \hbbc\ .

Similarly, the linearized hydrodynamics predicts the existence of a simple pole
in the correlators of the (longitudinal) components  of  $R$-currents, with
the dispersion relation
\eqn\hbgs{\omega = -  i D_R \, q^2\,,}
where $D_R$ is the $R$-charge diffusion constant.

One should keep in mind that the dispersion relations above
are valid in the domain of long wavelengths
 and will generically have corrections
containing higher powers of $q$.

The regime of finite-temperature LST accessible to supergravity and
tree level string theory calculations is the theory at the Hagedorn  temperature.
From thermodynamics it follows that the speed of sound vanishes at $T=T_H$.
Moreover, universality results for the shear viscosity obtained
from supergravity  \refs{\KSSa,\KSSb\BuchelTZ-\BuchelQQ},
suggest that
 the ratio $\eta/s$, where $s=S/V_5$ is the entropy density,
  remains finite and equal to  $1/4\pi$ at $T=T_H$,
 at least in
the supergravity approximation. Then, since $\epsilon + P = s T$,
 knowing the sound attenuation constant \hbbxc\ allows one
to compute the ratio of bulk viscosity to entropy density.

In the remaining part
 of the paper we compute the retarded Green's functions
of the stress-tensor and $R$-current correlators and analyze their singularities.
The poles computed in supergravity agree
with the string theory results exactly, and do not receive $1/k$
corrections.
These results also agree with the predictions of the
linearized hydrodynamics.

In summary, we find that

$\bullet$
 The shear mode correlators  have a simple pole predicted by \hbbs\ ,
with $\eta/s=1/4\pi$.

$\bullet$
 The scalar mode correlators do not have hydrodynamic poles.

$\bullet$
 The $T^{xx}$ mode in the sound channel decouples, and thus according to
\rgfs\ we have
 \eqn\rgfsa{v_s = 0 \,, \;\;\; {\zeta\over s} =
 {2 \over 5}\, {\eta\over s} = {1\over 10 \pi}\,.}

$\bullet$ Correlators of the sound modes exhibit a {\it double}
pole at $\omega = - i q^2/4\pi T$. 
One is tempted to view it either as
merging of two simple poles $(\omega - |v_s| q + i \Gamma q^2)
 (\omega + |v_s| q + i \Gamma q^2)$ in the limit $v_s\rightarrow 0$,
or, ignoring $q^4$ terms unaccounted for in linearized hydrodynamics,
as a simple pole \hbbsx\ . Each interpretation leads to the same
attenuation constant, which gives $\zeta /s = 1/10\pi$ 
coinciding with 
\rgfsa\ . However, such an interpretation is problematic: at $v_s$ strictly
 zero,  solutions to the dispersion equation 
\hbgs\ are given by $\omega =0$ and $\omega = -i\Gamma q^2$ rather than by a 
double root at $\omega = - i \Gamma q^2/2$. At the same time, 
introducing quartic terms into the hydrodynamic equations 
requires further analysis.

$\bullet$
 Correlators of the longitudinal components of $R$-currents have a simple
pole at $\omega$ given by \hbgs\ with the diffusion constant $D_R = 1/4\pi T$.

$\bullet$ These results are exact to all orders in $1/k$ (
or equivalently, to all orders in $\alpha'$).
\newsec{Details of the world-sheet description}
We consider a system of $k$ non-extremal NS5 branes.
The Euclidean version of the near horizon geometry defines an exact superconformal
field theory $\IR^5 \times  {SL(2)\over U(1)}\times SU(2)$.
We denote by $X^\mu$ coordinates on $\IR^5$ and by $\psi_\mu$ their
superpartners.

Here we summarize some useful facts on supersymmetric $SL(2)/U(1)$
at level $k$.
We set $\alpha'=2$.
The semiclassical geometry of Euclidean  ${SL(2)\over U(1)}$ is that of
a cigar
\eqn\mcigar{
\eqalign{  ds^2&=2 k\left(d\phi^2+\tanh^2\phi d\tau^2\right)\,,\cr
            \Phi&=\Phi_0-\log\cosh\phi \,.  \cr}
}
Here $\Phi_0$ is the value of the dilaton at the tip of the cigar.
Far from the tip, the background has an asymptotic form
of a cylinder with linear dilaton.
Both $\phi$ and $\tau$ have their fermion superpartners $\psi_\phi$ and
$\psi_\tau$.
The central charge of the cigar theory is $c_{SL(2)/U(1)}=3+6/k$,
so that the total central charge is $15/2+3+6/k+9/2-6/k=15$.

Below we focus on the quantities which are holomorphic on the worldsheet
(there are similar expressions for their antiholomorphic counterparts).
The asymptotic expressions for the  generators of the
$\NN=2$ worldsheet superconformal algebra can be found in e.g.  \AFKS\ , \GKPS :
\eqn\ntwoalg{ G^+=i\psi\p X^*+i Q \p\psi,\qquad  G^-=i\psi^*\p X+i Q \p\psi^*,\qquad
                        J=\psi\psi^*+iQ\p\tau\,,}
where $ \psi= (\psi_\phi+i\psi_\tau)/\sqrt{2}$ and $Q=\sqrt{2/k}$.
The important set of observables in the $SL(2)/U(1)$ model
consists of Virasoro primaries $V_{jm}$ with the
conformal dimension and the $U(1)_R$ charge given respectively by
\eqn\dimnq{\Delta[V_{jm}]=-{j(j+1)\over k}+{m^2\over k},\qquad   R[V_{jm}]={2m\over k} \,. }
The asymptotic behavior of $V_{jm}$ is
\eqn\vasymp{  V_{jm}\cong { e^{i m Q \tau} e^{j Q \phi}\over 2j+1} \,.  }
This allows us to compute the action of superconformal generators on $V_{jm}$:
\eqn\gactv{  G^+_{-{1\over 2}} V_{jm}\cong -iQ (j+m) \psi V_{jm},\qquad
           G^-_{-{1\over 2}} V_{jm}\cong -iQ (j-m) \psi^* V_{jm}\,.  }

The supersymmetric $SU(2)_k$ ($k$ here defines the level) can be
decomposed into the bosonic $SU(2)_{k-2}$ with $SU(2)$ currents
$J^{A},\;A=1,2,3$ and free fermions $\psi^A$ with an OPE
\eqn\opferm{  \psi^A(z_1) \psi^B(z_2)\sim {\delta^{AB}\over z_1-z_2} \,. }
The $SU(2)$ currents of the supersymmetric model are given by
\eqn\susycurrents{ J^{A,susy}=J^A-i\epsilon^A_{\;BC}\psi^B\psi^C
\,. }

\newsec{Two-point function of the stress-energy tensor}
Here we compute the two-point function of the stress-energy tensor
\rgf. According to the holographic prescription, this problem is
equivalent to computing the two-point function of the graviton in
the dual string theory. Since we are interested in the pole
structure, we will neglect an overall normalization
coefficient\foot{This normalization coefficient diverges
exponentially at high momenta, signifying the non-locality of LST
\refs{\PeetWN\MinwallaXI-\KapustinCI}. It approaches a constant in
the hydrodynamic regime and does not affect the poles.}. According
to \rgf, the graviton has energy $\omega$ and spatial momentum $q$
which is aligned along the $z$ direction. The polarization has one
leg along $z$ and one leg along $x^a$.
String theory computation is performed in Euclidean space, making $\omega$
quantized
in the units of temperature.
To recover the Lorentzian version of the correlator, we must perform
analytic continuation to imaginary frequencies.
\subsec{Transverse polarization}
We first review the computation for transversely polarized graviton \refs{\GKa,\GKb}.
Moreover, in
\NarayanDR\ the string theory result was compared with
the one obtained in (Euclidean) supergravity, finding
 agreement up to the terms suppressed by $1/k$
(see also \DeBoerDD).
The matter part of the transverse graviton vertex operator in the (-1,-1) picture is
\eqn\tgv{ V^t= c {\bar c}e^{-{\varphi}-{\bar\varphi}}
                \xi^{ab}\psi_a{\bar\psi_b} e^{i q z} V_{jm\bar m}\,.  }
Here
$\xi^{ab}=\xi^{ba}$ is the polarization tensor,
${\varphi}$ and ${\bar\varphi}$ are (anti)holomorphic
superconformal ghosts, and $\psi_a$ and $\bar\psi_a$ are
(anti)holomorphic fermionic superpartners of the transverse
coordinates on the five-brane worldvolume $x^a\neq z$,
$a=1,\ldots,4$, $V_{jm\bar m}$ is the primary of the $\NN=2$
superconformal algebra of $SU(2)/U(1)$. We consider the case of
vanishing winding number, thus ${\bar m}=-m$.
The GSO projection implies $m\in k \IZ$.
Physical state condition relates $j$ with $m$ and $q$:
\eqn\physstate{ -{j(j+1)\over k}+{m^2\over k}+{q^2\over 2}=0 \,. }
One can now solve for $j$. The holographic prescription implies
that $j$ must correspond to the state which is not normalizable
in the cigar.
The condition  
of non-normalizability $j>1/2$ \refs{\GKa,\MaldacenaHW} imposes the choice
of sign of the square root:
\eqn\jmq{   j=-{1\over 2}+{\sqrt{1+4 m^2 +2 k q^2}\over 2} \,.  }
The two-point function can be read from \refs{\GKa,\GKb}:
\eqn\twograv{   \Pi(j,m)=
                        {\Gamma(1-{2j+1\over k})\Gamma(-2j-1) \Gamma^2(j-m+1)
\over\Gamma({2j+1\over k})\Gamma(2j+1)\Gamma^2(-j-m)}\,.   }
Note that this formula is invariant under $m\ra -m$, as
long as $m\in \IZ$, which is the case here.
To compare with supergravity, we must identify
parameters in the following way
\eqn\map{  T={1\over 2\pi \sqrt{2 k}},\qquad \omega =-{m\over \sqrt{2k}}\,.  }
It is also useful to define
\eqn\defbold{  \w_E= {\omega \over 2\pi T}=-2 m,\qquad  \q={q\over 2\pi T} \,. }
Hence, \jmq\ can be re-casted as
\eqn\jmqa{ j=-{1\over 2}+{\sqrt{1+\w_E^2 +\q^2}\over 2}\,.   }
Now we can rewrite the two-point function of transverse graviton in
the form it appears in \NarayanDR\
\eqn\twograva{   \Pi(\q,\w_E)=
 {\Gamma\left(1-{\sqrt{1+\w_E^2 +\q^2}\over 2k}\right)\Gamma\left(-\sqrt{1+\w_E^2 +\q^2}\right)
 \Gamma^2\left({1+\w_E\over 2}+{\sqrt{1+\w_E^2 +\q^2}\over 2}\right)\over
 \Gamma\left({\sqrt{1+\w_E^2 +\q^2}\over 2k}\right)
 \Gamma\left(1+\sqrt{1+\w_E^2 +\q^2}\right)
 \Gamma^2\left({1+\w_E\over 2}-{\sqrt{1+\w_E^2 +\q^2}\over 2}\right)
}  \,. }
This formula, except for the first factor, has been also computed
in supergravity \NarayanDR.
To obtain retarded Green's function of the transverse components of
the stress-energy tensor, \twograva\ must be analytically continued
to Minkowski space.
Substitution $\w_E= -i \w$ brings it to the form
\eqn\rgftr{  G_{x^ax^b,x^ax^b}(\q,\w)\sim
    {\Gamma\left(1-{\sqrt{1-\w^2 +\q^2}\over 2k}\right)\Gamma\left(-\sqrt{1-\w^2 +\q^2}\right)
 \Gamma^2\left({1-i\w\over 2}+{\sqrt{1-\w^2 +\q^2}\over 2}\right)\over
 \Gamma\left({\sqrt{1-\w^2 +\q^2}\over 2k}\right)
 \Gamma\left(1+\sqrt{1-\w^2 +\q^2}\right)
 \Gamma^2\left({1-i\w\over 2}-{\sqrt{1-\w^2 +\q^2}\over 2}\right)
} \,.}
This formula also appears in \DeBoerDD.

\subsec{Longitudinal polarization}
Having completed  the exercise with the transverse graviton, let us
consider polarization that is longitudinal on the boundary. The
vertex operator has the following asymptotic form \eqn\lpg{
V^{l}=c{\bar c}e^{-{\varphi}-{\bar\varphi}}
   \xi^{za}\left[(\psi_z+A \psi_\phi){\bar\psi_a}+\psi_a ({\bar\psi}_z+A {\bar\psi}_\phi)\right]
    e^{i q z} V_{jm\bar m}   \,. }
For a moment we will concentrate on the holomorphic part of
the vertex operator,
\eqn\hollg{  (\psi_z+A \psi_\phi) e^{i qz} V_{jm} \,. }
We must also require \lpg\ to be BRST-invariant.
That is, \hollg\ must be annihilated by the action of $(L_0-{1\over2})$ and $G_{1/2}$.
The former condition leads to \jmq.
The latter determines $A$, as we show momentarily.
We can make use of \gactv\ to rewrite \hollg\ as
\eqn\hollga{e^{iqz} \left(\psi_z+
           A({1\over j+m}G^+_{-{1\over2}} + {1\over j-m}G^-_{-{1\over2}} )\right)
     V_{jm}\,.  }
Acting by $G_{1/2}=(G^+_{1/2}  +G^-_{1/2})/\sqrt{2}$ we deduce
\eqn\valuea{
 A= -{\sqrt{2} q (j^2-m^2)\over {4m^2\over k}-2 j q^2}\,.}
In the derivation of \valuea\ we used the $\NN=2$ superconformal algebra together with
\eqn\lzerov{ L_0 V = \Delta[V] V,\qquad \Delta[V]=-{q^2\over 2} \,, }
and
\eqn\qzerov{  J_0 V={2 m\over k} V \,. }
To summarize, \hollg\ can be written as
\eqn\hollgb{   e^{i qz} \left(\psi_z-{\sqrt{2} q\over{4m^2\over k}-2 j q^2}
        \left[(j-m)G^+_{-{1\over 2}}+(j+m)G^-_{-{1\over 2}}\right]\right) V_{jm} \,. }
In computing the two-point correlator $\langle V^{az}(z)V^{az}(0)\rangle$
 the following identity will be useful
\eqn\ida{
\eqalign{
  &\langle \left[(j-m)G^+_{-{1\over 2}}+(j+m)G^-_{-{1\over 2}}\right] V(z_1)
          \left[(j-m)G^+_{-{1\over 2}}+(j+m)G^-_{-{1\over 2}}\right] V(z_2)\rangle = \cr
  &\qquad -2 (j^2-m^2) \langle L_{-1} V(z_1) V(z_2)\rangle=
                 -2(j^2-m^2)q^2 z^{-1} \langle V(z_1) V(z_2)\rangle \,,  \cr}
}
where we used \lzerov.
Eqs. \hollga, \valuea, and \ida\ allow us to compute the two-point function
of the graviton that is longitudinally polarized on the boundary
\eqn\twogravl{  \left[1-{4 q^4 (j^2-m^2)\over
             \left({4 m^2\over k}-2 j q^2\right)^2}\right] \Pi(\q,\w_E)\,,  }
where $\Pi(\q,\w_E)$ and $j$ are given by \twograva\ and  \jmqa\ , respectively.
Eq.~\twogravl\ can be re-casted as
\eqn\twogravla{  {\w_E^2\left(\w_E^2-2 j \q^2+{\q^4\over 4}\right)\over
        \left(\w_E^2-j\q^2\right)^2} \Pi(\q,\w_E) \,. }
Performing analytic continuation, we obtain the expression for the two-point
function corresponding to the shear mode
\eqn\shearg{
  G_{x^az,x^az}(\q,\w)\sim
    {\w^2\left(\w^2+2 j \q^2-{\q^4\over 4}\right)\over \left(\w^2+j\q^2\right)^2}
  {\Gamma\left(1-{2j+1\over k}\right)\Gamma\left(-2j-1\right)
 \Gamma^2\left(1+i{\w\over 2}+j\right)\over
 \Gamma\left({2j+1\over k}\right)\Gamma\left(2j+2\right)
 \Gamma^2\left(-i{\w\over 2}-j\right)
}\,,
}
where
\eqn\jwqm{  j=-{1\over 2}+{\sqrt{1-\w^2+\q^2}\over 2}\,.        }
The Green's function for the sound mode is computed in a similar
manner.
One simply needs to notice that both holomorphic and antiholomorphic
parts of the vertex operator take the form of \hollg.
The result for the Green's function is then
\eqn\soundg{
  G_{zz,zz}(\q,\w)\sim
    \left[{\w^2\left(\w^2+2 j \q^2-{\q^4\over 4}\right)\over \left(\w^2+j\q^2\right)^2}\right]^2
  {\Gamma\left(1-{2j+1\over k}\right)\Gamma\left(-2j-1\right)
 \Gamma^2\left(1+i{\w\over 2}+j\right)\over
 \Gamma\left({2j+1\over k}\right)\Gamma\left(2j+2\right)
 \Gamma^2\left(-i{\w\over 2}-j\right)
}\,.
}
\subsec{R-charge diffusion constant}
The vertex operator dual to the transverse component
of the $SU(2)_R$ current in LST $J^{lst,\;  B}_{x^a}$ is
\eqn\voprct{  V^{t,\;B}=c {\bar c}e^{-{\varphi}-{\bar\varphi}}
                \left[\psi^B{\bar\psi_a}+\psi_a {\bar \psi}^B\right] e^{i q z} V_{jm\bar m}\,.  }
The two-point function is computed as in section 5.1.
The result is
\eqn\rdt{  G^{BC}_{x^ax^b}(\q,\w)\sim \delta^{BC} \delta_{ab}
{\Gamma\left(1-{\sqrt{1-\w^2 +\q^2}\over 2k}\right)\Gamma\left(-\sqrt{1-\w^2 +\q^2}\right)
 \Gamma^2\left({1-i\w\over 2}+{\sqrt{1-\w^2 +\q^2}\over 2}\right)\over
 \Gamma\left({\sqrt{1-\w^2 +\q^2}\over 2k}\right)
 \Gamma\left(1+\sqrt{1-\w^2 +\q^2}\right)
 \Gamma^2\left({1-i\w\over 2}-{\sqrt{1-\w^2 +\q^2}\over 2}\right)
} \,.}
The vertex operator dual to the longitudinal component
of the $SU(2)_R$ current in LST $J^{lst,\;  B}_{z}$ is
\eqn\voprcl{  V^{l,\;B}=c{\bar c}e^{-{\varphi}-{\bar\varphi}}
   \left[(\psi_z+A \psi_\phi){\bar\psi^B}+\psi^B({\bar\psi}_z+A {\bar\psi}_\phi)\right]
    e^{i q z} V_{jm\bar m}   \,. }
and the retarded Green's function for $J^{lst,\;  B}_{z}$ is
\eqn\rdl{
  G^{BC}_{zz}(\q,\w)\sim \delta^{BC}
    {\w^2\left(\w^2+2 j \q^2-{\q^4\over 4}\right)\over \left(\w^2+j\q^2\right)^2}
  {\Gamma\left(1-{2j+1\over k}\right)\Gamma\left(-2j-1\right)
 \Gamma^2\left(1+i{\w\over 2}+j\right)\over
 \Gamma\left({2j+1\over k}\right)\Gamma\left(2j+2\right)
 \Gamma^2\left(-i{\w\over 2}-j\right)
}\,.
}

\newsec{The poles of the correlators and their interpretation}
We will be
mostly interested in the poles of the Green's functions which
correspond to the excitations without a gap, i.e. the hydrodynamic
poles with the property
 $\w\ra 0$ as $\q\ra 0$.
In this limit $j\ra 0$.
Consider first the shear mode [eq.  \shearg].
A possible source of poles is the denominator
$\left(\w^2+j\q^2\right)^2$. The equation \eqn\wjq{   \w^2+j\q^2=0
} has a simple solution  $\w^2=-\q^4/ 4$,  $j=\q^2/ 4$.
Hence the denominator appears to contribute
two double poles at
\eqn\poleb{   \w=\pm i{\q^2\over 2}    \,.         }
However,  the numerator in the first factor has a simple zero at \poleb
\eqn\zeron{  \left(\w^2+2 j \q^2-{\q^4\over 4}\right)=0\,\,\;\;\;  {\rm for}\,\,\;\;\;
 \w^2=-{\q^4\over 4} \,.  }
Hence  the first factor in \shearg\ contributes only two single poles
at $\w$ given by \poleb.
One of these poles is cancelled by a zero coming from $\Gamma^{-2}\left(-i{\w\over 2}-j\right)$.
Indeed, \poleb\ with a plus sign is a solution of
\eqn\wjeq{  -i{\w\over 2}-j=0\,. }
Therefore we are left with a single hydrodynamic pole at $\w = - i \q^2/2$.
In addition, there are gapless poles at $\w=\pm \q$ coming
from $\Gamma\left(-\sqrt{1-\w^2 +\q^2}\right)$.

To summarize, the retarded Green's function for the shear mode
has the form
\eqn\gfpoles{   G_{x^az,x^az}\sim {1\over (\w+\q)(\w-\q)(i\w-{\q^2\over 2})   } \,,  }
where we only exhibit the structure of poles which correspond to excitations
without a gap.
In addition to the poles that correspond to the propagating modes, there
is a single hydrodynamic pole at
\eqn\shearpole{\omega = -i q^2/ 4 \pi T\,.}
Comparing with Eq.~\hbbs\ we find  $\eta/s=1/4\pi$.

Turning to the correlators in the sound channel, we observe that
the difference between Eq.~\shearg\ and Eq.~\soundg\ is that in
Eq.~\soundg\ the prefactor is squared. We immediately conclude
that in the hydrodynamic regime the correlator $G_{zz,zz}$ has the
form \eqn\gfpoless{   G_{zz,zz}\sim
 {1\over (\w+\q)(\w-\q)(i\w-{\q^2\over 2})^2   } \,.  }
Comparing this to the discussion in Section 3 we find the speed of sound and
 the ratio of bulk viscosity to entropy density at $T=T_H$:
\eqn\zvuk{v_s=0\,, \;\;\;\;\;\qquad \;\;\;\;  {\zeta\over s} =
1/10\pi\,.} Note that these results are exact to all orders in
$1/k$.

The pole structure of the Green's functions for the R-currents is
analyzed in a similar manner. It is sufficient to note that \rdt\
is proportional to \rgftr\ and \rdl\ is proportional to \shearg.
That is, \eqn\rchdpoles{ G^{BC}_{zz}\sim {\delta^{BC}  \over
(\w+\q)(\w-\q)(i\w- {\q^2\over2}) } \,.} Comparing with \hbgs\ we
find the value of the R-charge diffusion constant to be
$D_R=1/(4\pi T_H)$.

There are also other poles, coming from $\Pi (\w,\q)$. These poles
are identical for all correlators, since all the correlators
contain the factor  $\Pi (\w,\q)$. The poles are given by
\eqn\zosx{\w = \pm \sqrt{ \q^2 + 1 - n^2}\,, \;\;\; n=1,2,...\,.}
 Note that the poles $\w = \pm \q$ (given by Eq.~\zosx\ with
 $n=1$) correspond to
  a mode propagating with the speed of light on the five-branes.
\ifig\frequ{Distribution of poles in the complex $\w$ plane for
$\q=1$. The hydrodynamic pole at $\w = - i \q^2/2$ is encircled.
This pole is absent for the scalar mode correlators. It is a
simple pole for
 the shear mode,
 and a double pole for the sound mode. All other poles are given
 by Eq.~\zosx\ . }
{\epsfxsize=9.5cm \epsfysize=5.5cm \epsfbox{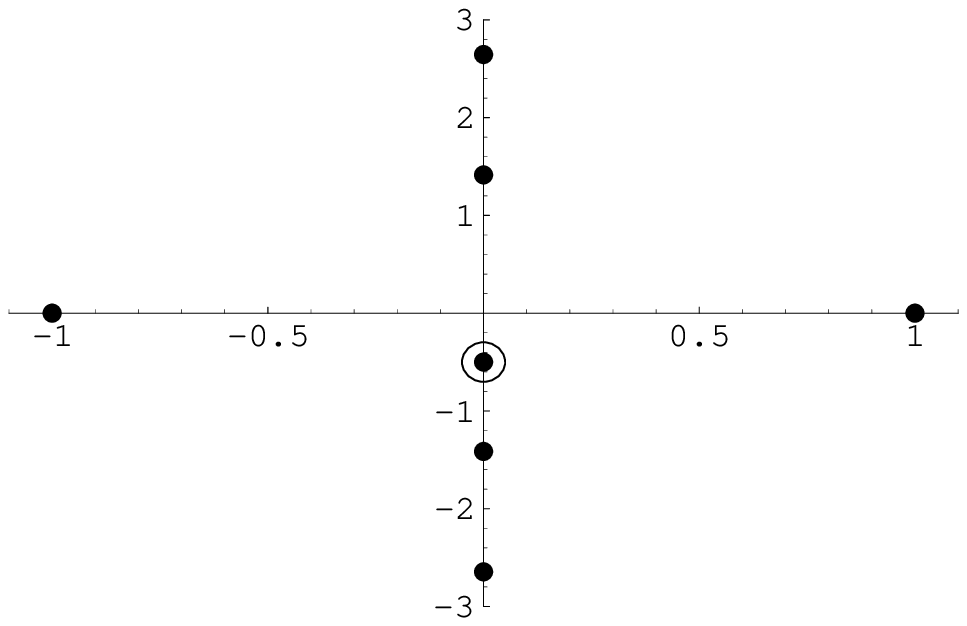}} This
mode does not have any usual field-theoretic interpretation, since
in thermal field theory one cannot have propagation without
attenuation. Interpretation of the poles with a finite gap in
Eq.~\zosx\ is even more problematic. For any fixed $n>1$ and
sufficiently large $q$, there is a pair of poles on the real axis.
In the limit $\q\rightarrow \infty$ there is an infinite number of
such poles accumulating on the real axis. At finite $q$, there are
also poles distributed symmetrically along the negative and
positive imaginary axis. This is incompatible with the basic
analyticity property of the retarded Green's function and perhaps
is a signal of an instability in the system. Yet another set of
poles arises from the gamma-function $\Gamma (1-(2j+1)/k)$. These
poles scale as $k$ for large $k$, $\w \sim \pm i k (n+1)$,
$n=0,1,2,...$
 and are not visible in supergravity approximation.
We shall return to the question of interpretation of the finite
gap poles as well as the massless pole in Section 8.

In the next Section, we confirm the results of the string
calculation by computing quasinormal spectra of non-extremal
NS5-branes in supergravity.

\newsec{Correlation functions from gravity}

For calculations in supergravity, it will be convenient to introduce
the new radial coordinate $u=r_0^2/r^2$. The background \nensfive\ becomes
%
\eqn\nonsfive{ ds^2=- f(u) dt^2 + dx_5^2
   + {r_0^2 A(u)\over u}
    \left({du^2\over 4 u^2 f(u)} + d\Omega_3^2\right)\,,}
\eqn\ldil{ e^{2\Phi}=g_s^2 A(u)\,,} \eqn\form{ H_3 = 2 L
\sqrt{L^2+r_0^2} \; \epsilon_3\,,}
 \eqn\newf{f(u) = 1- u\,,}
  \eqn\newa{ A(u) =  1 + {L^2 u \over r_0^2}\,,}
  where $L= k \alpha'$.
Explicitly, we use the coordinates $\phi_1$, $\phi_2$, $\phi_3$ on
the sphere, with \eqn\newo{d \Omega_3^2 = d\phi_1^2 + \sin^2
\phi_1 d \phi_2^2 + \sin^2\phi_1 \sin^2\phi_2 d\phi_3^2\,, }
\eqn\vfo{\epsilon_3 = \sin^2 \phi_1 \sin \phi_2 \; d \phi_1 \wedge d \phi_2
\wedge d \phi_3\,.}

The background \nonsfive\ ---\form\ is a solution to the type II
supergravity equations of motion \eqn\newra{R_{\mu\nu} = - 2
\nabla_{\mu}\partial_{\nu} \Phi + {1\over 4} H_{\mu \alpha \beta}
H_{\nu}^{\alpha\beta}\,,} \eqn\newdi{\nabla^2 \Phi =
\partial_{\mu} \Phi
\partial^{\mu}\Phi + {1\over 48}\, H^2_3 - {1\over 4} \, R\,,}
\eqn\newho{d * \;  H_3 e^{-2\Phi} =0\,,} with all other
supergravity fields consistently set to zero.

The near-horizon limit $r_0/L \rightarrow 0$ of the NS5 brane
background  \nonsfive\ - \form\ provides an effective
description of LST at high energies.

It will be convenient to choose the spatial momentum along one of
the coordinate directions on the brane. In the following we use
$z$ to denote the coordinate along which the momentum is directed.

Fluctuations\foot{The notation $\varphi$ was used earlier in the paper 
to denote the superconformal ghost field. Here and henceforth 
we use the same notation to denote dilaton's fluctuation. 
We hope this will not lead to a confusion.}
 $\delta g_{\mu\nu} \equiv
h_{\mu\nu}(u,t,z)$, $\delta \Phi \equiv \varphi (u,t,z)$
of the background \nonsfive\ fall into three
categories\foot{Fluctuations of the three-form field can be consistently
 set to zero. Eq.~\newho\ is automatically satisfied for fluctuations
 independent of the angular coordinates.}
 corresponding to the
 scalar, shear and sound mode channels of the stress-tensor
correlation function \PSS, \KST\ :
\eqn\mch{   Scalar \;\; mode: \; \;  H = h_{x^a x^b}\,,\;\;\;
a\neq b, \;\;\; a,b=1,...4\,
}
\eqn\mch{   Shear \;\; mode: \; \;
 H_{tx} = h_{t x^a}\,, \;\;\;
 H_{zx} = h_{z x^a}\,, \;\;\; \forall a=1,...4\,
}
\eqn\mch{    Sound \;\; mode: \; \;  \varphi\,, \;\; H_{tt} = h_{tt}/f\,, \;\;
   H_{tz} = h_{tz}\,, \;\;   H_{zz} =  h_{zz}\,,  \;\;
H_{xx} = \sum\limits_{a=1}^4  h_{x^a x^a}\,.
}
Fluctuation equations for each
of these modes decouple, and can be considered separately.

In addition, a
 convenient way of dealing with the fluctuation equations is to introduce
variables invariant under the infinitesimal diffeomorphisms \KST\
\eqn\nqq{\eqalign{x^{\mu} &\rightarrow x^{\mu} + \xi^{\mu}\,,
\cr
g_{\mu\nu} &\rightarrow g_{\mu\nu} - \nabla_{\mu} \xi_{\nu} -  \nabla_{\nu}
 \xi_{\mu}\,,
\cr
\varphi  &\rightarrow \varphi - \partial^{\mu} \Phi\,  \xi_{\mu}\,.\cr
}
}
Assuming the dependence of all fields
on $t$ and $z$ to be of the form $\propto e^{-i \omega t + i q z}$,
one identifies the following gauge-invariant variables for the three
channels:
\eqn\mch{   Scalar \;\; mode: \; \;  Z_2 = H \,,
}
\eqn\mch{   Shear \;\; mode: \; \; Z_1 = q H_{tx} + \omega H_{zx}\,,
}
\eqn\mch{    Sound \;\; mode: \; \; Z_h = q^2 f H_{tt} + 2 \omega q H_{tz} + \omega^2 H_{zz} - 2 u q^2 \varphi\,,  \;\; Z_{\varphi} = H_{xx}\,.
}

\subsec{ The scalar mode}
 For the scalar mode $Z_2(u)$, the only nontrivial
equation coming from the system \newra\ - \newho\ in the
near-horizon limit is \eqn\nqqs{ Z_2'' - {1\over f} Z_2 ' + { \w^2
- \q^2 f\over 4 u^2 f^2} Z_2 =0\,, } where  $\w = \omega/ 2 \pi
T_H$, $\q = q/ 2 \pi T_H$. Here $T_H = 1/2\pi L$ is the
near-horizon limit of the Hawking temperature associated with the
metric \nonsfive\ .
 Eq.~\nqqs\ is a hypergeometric equation whose solution obeying the incoming
wave boundary condition at the horizon $u=1$ is
\eqn\niqss{Z_2(u) = C (1-u)^{-i {\w\over 2}} u^{\varrho}
 \ofo \left( - {i \w\over 2} +\varrho\,,  - {i \w\over 2} +\varrho\,;\,
 1-i\w;\, 1-u\right)\,,}
where $C$ is the normalization constant,
\eqn\nixs{\varrho = {1\over 2} \left( 1 - \sqrt{1 + \q^2 - \w^2}\right)\,.
}
In the limit $u\rightarrow 0$ the asymptotics of the solution
\niqss\ is
\eqn\niaqss{ Z_2 (u) \sim {\cal A} \, u^{\varrho} +\cdots +
 {\cal B} \, u^{1-2\varrho}
 +\cdots\,,
}
where ${\cal A}$,  ${\cal B}$ are the coefficients of the connection matrix
of the hypergeometric equation. The location of the
poles of the retarded correlation
 function corresponding to the perturbation $H$ can be found by imposing
a Dirichlet boundary condition
\eqn\zos{ Z_2 (0) =  {\cal A}  = { \Gamma ( 1-i \w)
 \Gamma ( \sqrt{1 + \q^2 - \w^2 })\over \Gamma^2 ( 1- i \w/2 - \varrho )} =0\,.
}
Eq.~\zos\ has no solutions for real $\q$.
Additional poles arise from the (apparent) singularities of the local solutions
at $u=0$, as explained in \KST\ . These are given by Eq.~\zosx\ .
Simple poles \zosx\ are precisely the
singularities of the correlator \rgftr\ .

In  the hydrodynamic regime $\w\ll 1$, $\q \ll 1$,
a perturbative solution to Eq.~\nqqs\ is given by
\eqn\niqs{ Z_2 (u) = C\, f^{-i \w/2}\, \left( 1 - {\w^2 \over 4} \, {\rm Li}_2 (1-u)
+ {\w^2 - \q^2\over 4} \log{u} \right) + \cdots\,,
}
where $C$ is (another) normalization constant, and ellipses denote terms
of higher order in $\w$, $\q$.

\subsec{The shear mode}
 The shear mode fluctuations $H_{tx}$,
$H_{zx}$ obey the system of equations obtained from
Eq.~\newra\
 \eqn\eoga{ \w H_{tx}' + \q f H_{zx}' =0\,,}
 \eqn\eogb{
 H_{tt}'' - {1\over 4 f u^2} \left( \w\q H_{zx} + \q^2 H_{tx}\right) =0\,,}
 \eqn\eogc{ H_{zx}'' - {1\over f} H_{zx}' + {1\over 4 u^2 f^2} \left(
\w^2 H_{zx} + \w\q H_{tx} \right) =0\,.} Using \eoga\ - \eogc\ ,
for the gauge-invariant variable \mch\ one finds \eqn\znh{Z_1''-
{\w^2\over f(\w^2 - \q^2 f)} Z_1' + {(\w^2 - \q^2 f)\over 4 u^2
f^2}  Z_1 =0\,.}
 Eq.~\znh\ can be solved perturbatively in the hydrodynamic limit
$\w\ll 1$, $\q\ll 1$. Assuming first that $\w$ and $\q$ are of the
 same order,
 we obtain
\eqn\hs{Z_1 (u) = C\, f^{-{i \w \over 2}} \left( 1 + {i \q^2
f(u)\over 2 \w} + O(\w^2,\q^2,\w\q )\right)\,.} Quasinormal
spectrum is determined by imposing the Dirichlet condition
$Z_1(0)=0$. This gives the hydrodynamic dispersion relation
\eqn\hydd{\w = - i \q^2/2\,,} which is precisely the pole of the
correlator \shearg\ . One may object that the result \hydd\ is not
reliable, since it implies $\w \sim \q^2$, whereas the
perturbative expansion was based on the assumption  $\w \sim \q$.
To refine the argument, let us introduce a new parameter $\varsigma =
\w/\q$ and expand again, assuming $\varsigma \sim \q$. We get
\eqn\hyddz{Z_1 (u) = C\, f^{-{i \varsigma \q \over 2}} \left( f(u) - {2
i \varsigma\over q} + O (\varsigma)\right)\,.} The Dirichlet condition
$Z_1(0)=0$ then gives $\varsigma = - i \q/2$, in agreement with \hydd\ .
All other terms in \hyddz\ are of order $\varsigma$ or higher, and thus
the result \hydd\ is correct.

In fact, the full quasinormal spectrum can be determined exactly.
Combining Eqs.~\eoga\ and \eogb\ , we obtain the second-order ODE
for $H_{tx}'\equiv y(u)$, \eqn\somm{y'' +{2-3 u\over u f} y'
+{\w^2-f\q^2\over 4 u^2 f^2} y =0\,.}
 The solution of Eq.~\somm\ obeying the incoming wave
boundary condition at the horizon is given by
 \eqn\somsol{H_{tx}'(u)=C\,
f^{-i {\w\over 2}} u^{\varrho-1} \ofo \left(-{i \w\over 2}-\varrho -1 ,
-{i \w\over 2}-\varrho -1 ; 1-i\w ;  1-u \right)\,,} where  $C$ is
the normalization constant, and $\varrho$ is given by \nixs\ . Now,
Eq.~\eogb\ implies \eqn\imp{Z_1(u) = {4 f u^2\over \q}\,
H_{tx}''(u)\,.} The Dirichlet condition then reads \eqn\llp{Z_1(0)
= \lim\limits_{u\rightarrow 0} {4 f u^2\over \q} H_{tx}''(u)
=0\,.} Computing the limit we find that the condition \llp\ is
equivalent to
 \eqn\gamgam{{\Gamma(1-i \w)\Gamma(\sqrt{1+\q^2-\w^2})\over
\Gamma(2-\varrho-{i\w\over 2}) \Gamma(-\varrho-{i \w\over 2})} =0\,.} The
unique (for real $\q$) solution to Eq.~\gamgam\  is $\w=-i
\q^2/2$.

Additional singularities of the two-point function come from the
coefficients of the local Frobenius solution at $u=0$. They are the same
as in the scalar case, and are given by Eq.~\zosx\ .

\subsec{The sound mode}
Fluctuations of the sound wave mode are described by the
system of equations derived from Eqs.~\newra\ , \newdi\
\eqn\novvrga{\eqalign{ &H_{tt}'' - H_{zz}''- H_{xx}'' +
H_{\varphi}'' - {1\over f} \left( {3\over 2} H_{tt}' - H_{zz}' -
H_{xx}' + H_{\varphi}' \right) - {1\over 4 u^2 f^2} \Biggl[ \q^2 f
H_{tt} \cr &+ \w^2 H_{zz} + 2 \w \q H_{tz} +\left( \w^2 - f
\q^2\right) \left(H_{xx}- H_{\varphi}\right)
 \Biggr] =0\,,\cr} }
\eqn\novvrgb{\eqalign{
&
H_{tt}'' - {1\over 2 f}
\left(   3 H_{tt}' - H_{zz}' -  H_{xx}'+ H_{\varphi} ' \right)
-  { 1\over 4 f^2 u^2} \Biggl( \q^2 f H_{tt} + \w^2 H_{zz}  + 2 \w\q H_{tz}
\cr
&+\w^2 H_{xx} -  \w^2 H_{\varphi}  \Biggr) =0\,, \cr}}
\eqn\novvrgc{
H_{tz}'' +  {\w\q \over 4 f u^2}\, \left( H_{xx} - H_{\varphi} \right) =0\,,}
\eqn\novvrgd{
2 f \left( \w H_{zz}' + \q H_{tz}' + \w H_{xx}'-  \w H_{\varphi} '\right)
+ \w H_{zz} + 2 \q H_{tz} + \w H_{xx} -  \w  H_{\varphi} =0\,, }
\eqn\novvrge{
H_{xx}'' - {1\over f}\, H_{xx}' + {\w^2 - f \q^2 \over 4 f^2 u^2}\, 
H_{xx} =0\,,}
\eqn\novvrgf{
H_{zz}'' - {1\over f}\, H_{zz}' + { 1\over 4 f^2 u^2}\left[
 \q^2 f H_{tt} +  2 \w\q H_{tz}
\w^2 H_{zz}  - \q^2 f (H_{xx}- H_{\varphi})  \right] =0\,, }
\eqn\novvrgg{
2\q f \left( H_{tt}' - H_{xx}'+ H_{\varphi} '\right)
 + 2\w H_{tz}' - \q H_{tt} =0\,, }
\eqn\novvrgh{
H_{tt}'' - H_{zz}'' - H_{xx}''+ H_{\varphi} ''
- {2-3 u\over 2 u f} \left(  H_{zz}' + H_{xx}' - H_{\varphi} '\right)
+ {2-5 u\over 2 u f}  H_{tt}' = 0\,,}
%
%
%
where $H_{\varphi}=4\varphi$.

Turning to  equations for the gauge-invariant variables
$Z_h$, $Z_{\varphi}$, we find that  $Z_{\varphi}$ satisfies the equation for the
minimally coupled massless scalar \nqqs\ , whereas the equation for
$Z_h$ reads
\eqn\nouu{\eqalign{&Z_h'' - {2 \w^2 - \q^2 u \over f (2\w^2 - \q^2(2-u))} Z_h'
+ { 2 \w^4 + \w^2\q^2 (3u-4)+ \q^4 (u^2-3u+2) - 4 \q^2 u^2 f\over
4 u^2 f^2 (2\w^2 - \q^2(2-u))} Z_h \cr
 &+ {8\q^2(\w^2-\q^2)\over 2\w^2 - \q^2(2-u)}
 Z_{\varphi}' - {4 \w^2 \q^2 \over f   (2\w^2 - \q^2(2-u))}  Z_{\varphi} =0\,.\cr
}
}
Eq.~\nouu\ can be solved perturbatively in the hydrodynamic regime.
Since this is the sound wave mode, the standard dispersion relation would imply
$\w \sim \q$. Assuming such a scaling and imposing Dirichlet boundary condition
 on the perturbative solution of Eq.~\nouu\ , we find instead that
$\w\sim \q^2$, similar to the behavior of the shear mode. This is
of course precisely what we expect if the speed of sound vanishes.
Introducing again  $\varsigma = \w/\q \sim \q$, we obtain
\eqn\nouuw{Z_\varphi =  C_{\varphi}\, f^{- {i \varsigma q\over 2}}\left(
1 - {\q^2\over 4} \log{u} +  O(\varsigma^3) \right)\,. } \eqn\nouuw{Z_h
= C_h\, f^{- {i \varsigma q\over 2}} \left( u + {\q^2\over 4} u \log{u}
- {(2\varsigma + i \q)^2\over 2} f(u) + O(\varsigma^3) \right)\,, } where
$C_{\varphi}$, $C_{h}$ are the normalization constants.
 The
Dirichlet condition $Z_h(0)=0$
leads to a double zero at $\varsigma = -
i\q/2$. This is exactly
 the double pole of
 the correlator \soundg.
 In addition, a familiar set of singularities \zosx\ comes from
the coefficients of the local Frobenius solution at $u=0$.

\subsec{$R$-charge diffusion constant}
Diffusion of the $R$-charge in the high-temeperature phase
of LST can be considered along the lines of \KSSa\ , by solving
the Einstein-Maxwell equations in the hydrodynamic approximation.
The NS5-brane metric in the Einstein frame reads \eqn\nonde{ ds^2=
A^{-1/4}\left( - f(r) dt^2 + dx_5^2\right)
   + A^{3/4}
    \left({dr^2\over f(r)}+r^2d\Omega_3^2\right)\,,}
 \eqn\newf{f(r) = 1- { r_0^2 \over r^2 }\,,}
  \eqn\newa{ A(r) = 1 +
  {k l_s^2\over r^2} \equiv 1 + {L^2\over r^2}\,.}
(The metric \nonde\ is thus the same as the Einstein frame metric
for the D5 brane.) Using Eq.~(3.6) of  \KSSa\ , we find the
$R$-charge diffusion constant \eqn\nondes{ D_R = {1\over 4\pi
T_H}\,.} The result \nondes\ implies that the longitudinal
components of the $R$-current correlators in the high-temperature
phase of LST should have a simple pole at $\w = -i \q^2/2$. This
is indeed the case, as Eq.~\rdl\ shows.

\newsec{Discussion}

We have computed transport coefficients in Little String Theory at
Hagedorn temperature.
 Our result for the correlation function in the
sound channel is compatible with predictions of hydrodynamics up to the
terms
 quartic in spatial momentum. To account for those terms, one
 needs to improve the hydrodynamic description, possibly by
 including higher-derivative terms in the constitutive relation
 \hbbc.
It would certainly be interesting to extend the analysis to
temperatures other than the Hagedorn temperature. This would
correspond to including higher loop corrections in the string
 amplitudes.
%

In addition to the 
hydrodynamic poles, for
sufficiently large values of spatial momenta all the correlators
have an identical set of singularities, including the poles on the
real axis in the complex $\omega$-plane, and the poles on the
negative and positive imaginary axis\foot{These poles were
previously found in the scalar channel in \DeBoerDD\ .}. Also, one
of the poles in the correlators formally corresponds to a mode
propagating with the speed of light. Normally, retarded
correlators cannot have poles both in the lower and upper
half-planes in a stable system, and, moreover, one does not expect
a purely propagating mode to exist in a thermal medium. Since the
characteristic wavelength of the poles with finite gap is
$\sqrt{k} l_s$, it is conceivable that their existence is related
to the non-locality of LST.

Poles of similar nature arise in the correlators of LST in a
double scaling limit \refs{\AharonyXN,\GKa,\GKb}. Authors of
\AharonyXN\ observed massless poles that do not correspond to
physical states in the $U(1)^{k-1}$ super Yang-Mills theory which
naively is supposed to be a good description of LST at low
energies. From the world-sheet point of view, the relevant
correlators on the cigar are saturated in the bulk, far from the
tip. It has been argued in \AharonyXN\ that these poles appear due
to the UV/IR mixing, i.e. highly massive states do not decouple in
the infrared of LST. Massive poles, which are analogous to the
non-hydrodynamic poles described at the end of Section 6, were
found to be of similar origin \AharonyXN. These poles, coming from
non-locality and the UV/IR mixing should be distinguished from the
other, more conventional poles, which correspond to the
normalizable states at the tip of the cigar. These poles
correspond to physical states in LST.

The instability of the high-energy phase of LST appears to be
similar to that of a Schwarzschild black hole whose specific heat
also diverges to minus infinity as $E\ra\infty$.
We find it curious that the speed of sound in LST vanishes
precisely at Hagedorn temperature.
As we mentioned in the Introduction, in more conventional systems
such a behavior might be associated with a phase transition.

We found that the hydrodynamic pole does not receive $\alpha'$
(or, equivalently, $1/k$) corrections, and the ratio $\eta/s$ is
equal to the universal value $1/4\pi$.
This should be contrasted with the results of \BuchelDI, where
the curvature corrections to the near-extremal D3-branes were
investigated.
In \BuchelDI\ it was found that such corrections increase $\eta/s$.
In the system with RR flux, turning on $\alpha'$ corrections
is associated with departing from the infinite value of the t'Hooft
coupling in the dual gauge theory.
This fits well with the proposal of  \KSSa,\KSSb, that the viscosity bound should be
saturated in strongly coupled systems.
The case without RR flux studied in this paper appears to
be fundamentally different.
Indeed, there is no known way in which the theory on $k$ NS5 branes
becomes weakly coupled at large energy densities, even when $k$ is small.
It would be interesting to see what effect lowering energy density has on
the value of $\eta/s$.

In \GiveonZM\ a large class of LST vacua dual to string
theory compactified on a singular $n$-dimensional Calabi-Yau manifold was constructed.
The backgrounds considered in \GiveonZM\ have a general form
\eqn\bgnc{ \IR^{d-1,1}\times \IR_\phi \times \NN }
where $\IR_\phi$ describes the linear dilaton direction
and $\NN$ defines a superconformal theory whose detailed properties are
discussed in \GiveonZM.
In Eq.~\bgnc\ $d=10-2 n$.
Maximally supersymmetric LST in 5+1 dimensions
discussed in our paper corresponds to the Calabi-Yau being
a two-dimensional ALE space.
(In this case $\NN=SU(2)_k$).
Choosing the Calabi-Yau to be a singular three-fold can give rise to NS5 branes wrapping
various Riemann surfaces \KlemmBJ.
String theory calculations in our paper generalize straightforwardly
to these cases.
Indeed, introducing finite temperature to the system described by Eq.~\bgnc\
and performing Wick rotation, we end up with the background
\eqn\bgnct{ \IR^{d-1}\times {SL(2)_{k'}\over U(1)} \times \NN }
where $k'$ is determined by requiring \bgnct\ to be a consistent
background for superstring propagation (total central charge
of the worldsheet matter should be equal to 15).
The computations of the stress-energy Green's function in Section 5
do not involve the $\NN$ theory, and therefore are unaltered.
Hence, our computation of $\eta/s$ and $\zeta/s$ is valid for
a large class of LSTs.\foot{
In general, $k'$ does not need to be an integer.
When $d\geq 4$, $k'$ is bounded from below by $k'=1$, which
in $d=4$ corresponds to the Calabi-Yau being the singular conifold.
The holographic computation of Section 5 requires
$j$ in Eq.~\jmq\ to define non-normalizable state in the cigar.
In the hydrodynamic limit $j\ra 0$, which indeed corresponds
to the non-normalizable state, as long as $k'\geq 1$ \GKa,\MaldacenaHW.
The bound on $k'$ can be violated when $d=2$.}
One interesting class of four-dimensional LSTs
involves wrapping NS5 branes around Seiberg-Witten curve
at the Argyres-Douglas point \refs{\ArgyresJJ\ArgyresXN-\EguchiVU}.
At low energy, the theory on the fivebranes flows to
four dimensional $\NN=2$ SCFT \refs{\ArgyresJJ\ArgyresXN-\EguchiVU}.

\vskip 1.5 cm
\centerline{\bf Acknowledgments}

\vskip .2 cm

A.P. thanks P.~Kovtun, D.~Kutasov, G.~Moore and
D.~Sahakyan for discussions.
A.O.S. would like to thank M.~Paczuski, M.~Rangamani,
 S.~Ross,  M.~Rozali, and
especially  P.K.~Kovtun and D.T.~Son for discussions,
B.~Spivak for correspondence, and C.~Meusburger for 
comments on the manuscript.
We also thank the organizers of the ``QCD and String Theory'' workshop
at the KITP, UC Santa Barbara, where this work was initiated.
The work of A.P. is supported in part by  DOE grant DE-FG02-96ER40949.
Research at Perimeter Institute is supported in part by funds
from NSERC of Canada.



 \listrefs

\bye